\documentclass[10pt,conference]{IEEEtran}
\usepackage{hyperref}
\usepackage{epsfig}
\usepackage{graphicx}

\usepackage{epsfig,alltt,tabularx,amsmath}
\usepackage{times}
\usepackage{hyperref}
\usepackage{mystyle}
\usepackage{law}
\newtheorem{definition}{Definition}

\usepackage{epsfig}
\usepackage{wrapfig}
\input epsf

{

\newenvironment{respar}{
  \begin{list}{}{\topsep 0pt}
     \item[]
}{
  \end{list}}
\def\In{\begin{respar}}
\def\Un{\end{respar}}

\fboxsep=10pt
\fboxrule=1pt

\newcounter{saved}

\def\save{\saveitem{saved}\addtocounter{saved}{-1}}

\def\resume{\stepcounter{saved}\setitem{saved}}
{\begin{rules} \resume}%
{\save \end{rules}}

\pagestyle{plain}
\pagenumbering{arabic}
\title{An Approach to Modularization\\ of Distributed Systems
 }

\author{\IEEEauthorblockN{Nafftaly Minsky}
\IEEEauthorblockA{Computer Science Department\\
Rutgers University\\
New Brunswick NJ\\
Email: minsky@rutgers.edu}}

\begin{document}

\maketitle
\bibliographystyle{plain}


\begin{abstract}
 Modularization is an important architectural principle underlying many types
 of complex systems. It tends to tame the complexity of systems,
to facilitate their management, and to enhance their flexibility with respect
to evolution.
 In software, modularization has been
practiced and studied thoroughly in \emph{local}, i.e. non-distributed systems.
But very little attention has been paid so far to modularization in distributed
systems. This is, in part, because distributed systems are
inherently  modularized, in the sense that the
internals of each component of such a system is  inaccessible to other
components, thus satisfying the Parnas hiding principle. 
 It is, however, the thesis of this paper that there is much to be gained by
 being able to treat groups of distributed components as modules, called here
 \emph{distributed modules}. And that besides the conventional hiding principle,
 distributed modularization  should provide additional capabilities, 
which rarely, if ever, figure in  conventional modularized systems.
These capabilities include, but are not limited to:  the ability to impose constraints  on which kind of messages can be sent
    from a given distributed-module to its outside; and the ability to create AOP-like
    crosscutting modules.
This paper introduces a model of 
 \emph{modular distributed system}, or MDS, which satisfies such capabilities,
and which is implemented via the LGI middleware.
\end{abstract}

\s{Introduction}\label{intro}
 Modularization is an important architectural principle underlying many types
 of complex systems. It tends to tame the complexity of systems,
to facilitate their management, and to enhance their flexibility with respect
to evolution. Indeed, modularity is clearly manifested in biological systems,
and it has been employed in many types of physical artifacts.
 In software, modularization has been
practiced and studied thoroughly   in \emph{local}, i.e. non-distributed\footnote{We prefer the term ``local'' over the more common,
  but somewhat awkward, ``non-distributed''systems.}, systems.
And the constraints required for effective modularization in such systems is
usually supported by programming languages.
 
But very little attention has been paid so far to modularization in distributed
systems. This is, in part, because distributed systems are
inherently  modularized, in the sense that the
internals of each component of such a system is  inaccessible to other
components, thus satisfying the Parnas \cite{par72-2} hiding principle. 
 This may seem to be a sufficient degree of modularization.
And,  it may not be obvious what other kind of
modularization is relevant, or even  possible, in distributed systems. Indeed,
the literature seems to be silent on this issue.

 It is, however, the thesis of this paper that there is much to be gained by
 being able to treat groups of distributed components as modules, which we call 
 \emph{distributed modules} (or \emph{d-modules} for short). In an analogy to
 conventional modules, a d-module should be able to hide its ``internals,'' by
 which we mean that only specified component parts of a given module---i.e.,
 the distributed components that belong to this module---can be reached via a
 specified types of messages from the rest of the system. Moreover, we maintain
 that distributed modularization  should provide some additional capabilities, 
which rarely, if ever, figure in  conventional modularized systems.
These capabilities include:
(1) The ability to limit access to the internals of a given module to certain part
    of the system.
(2) The ability to impose constraints  on which kind of messages can be sent
    from a given d-module to its outside.
(3)  The ability of different modules to  overlap with each other.
 And (4) the ability to construct modules that implement AOP-like crosscutting
 concerns.

Such distributed modularization, we maintain, is particularly important for the
increasingly fragmented distributed systems, consisting of highly heterogeneous
components dispersed over the Internet, which may be written in different
languages, may run on different kinds platforms, and may be designed,
constructed, and even maintained under different administrative domains.  We
refer to such systems as having an \emph{open architecture}, or simply being
 \emph{open}\footnote{The term ``open
 system,'' as used here, has nothing to do 
 with the concept of \emph{open source}.}.
 For a prominent example of this class of systems
consider  the concept of \emph{service oriented architecture}
\cite{pap06-1} (SOA)---which is being adopted by a wide range of
complex distributed systems, such as enterprise systems,
federations such as \emph{grids},
 \emph{virtual organizations} (VOs), and 
\emph{supply chains}.

The question is how to formulate and enforce the constraints  on communication between system
components, which is required for establishing distributed modularization.
This cannot be done by means of  programming
language, as the various components of a given distributed system may be
written in many different languages. And it cannot be done by
controlling the code of the various components---particularly not in open systems---because one may have no access
to the code of many of them,  and no ability
to control such  codes. 

It is, however, possible to enforce the required constraints on communication
between components by means of a suitable middleware that is oblivious of the
code of the communicating components, and is thus independent of it.
Using such a middleware, we introduce in this paper
 a model of a \emph{modular
distributed system}, or MDS, which possesses the capabilities outlined above.
And we provide a proof of concept for this model via an experimental
implementation of it.

The rest of this paper is organized as follows:
 \secRef{motivation} motivates the need for distributed modularization
 via two examples.
 \secRef{requirements} introduces 
basic set of  requirements that an MDS, and its implementation, needs to satisfy to be effective.
\secRef{middle} discusses the type of middleware required for the
implementation of   our concept of
MDS, and describe the specific middleware   we use for this purpose.
\secRef{model} describes a basic model of MDS; and
 \secRef{advanced} extends this simplified model to  our complete MDS
model. Finally, \secRef{future} outlines some open problems by our model of
MDS;
 and we conclude in  \secRef{conclusion}.

\s{Motivating Examples}\label{motivation}
We consider here two examples of groups of components of  a distributed
system that provide a motivation for distributed modularization, and suggest
some of the properties it should have.

\noindent \textbf{A Monitoring Service:}
 Consider a  monitoring service ($MS$) designed for the management of large and
 geographically distributed system $S$.
The function of this service is to log various events occurring in $S$, analyze
them, and  provide reports about the results of this analysis to certain qualified
components of $S$; and to certain Internet cites outside of $S$, such as cites
that represent outside auditors.

 Suppose that $MS$ consists  of a group of components
 within $S$, partitioned into
 three disjoint sets: (1) the \emph{logs} that accept and maintains various types of logging
notices about events occurring in different parts of 
 system $S$; (2) the \emph{analyzers} that analyze the
various logs, for various purposes; and (c) the
\emph{reporters}, who prepare reports of various  results of this analysis for
stakeholders such as  auditors, managers, and  performance  optimizers---inside
the system, or outside of it.

 Suppose also that the APIs of these components provide the following
 operations, among others, intended for the use of clients outside of the MS group.
The logs feature the operations: (a) \TT{store}. which logs a given event, 
and (b) \TT{retrieve}, which gets some information from a log.
And the reporters feature and operation \TT{disclose} which sends to the
requester the type of reports it requires, if it is deemed to have the right
for it.

Now, it should be clear that the information handled by the MS group of
components could be very sensitive, depending on the nature of the system $S$
itself. So that the integrity and confidentiality of this information could be
critical and should be protected. One way to provide such protection is 
to ensure that the following
constraint on communication of components of MS with the rest of the Internet
are strictly enforced.

\begin{itemize}
\item The \TT{retrieve} operation of logs should not be visible  outside the
      $MS$ group. (This is because the raw data in logs may be very sensitive,
      and should not reveled, uninterpreted, to anybody outside of MS. It
      should, of course, be accessible to other components of MS itself.)
\item The \TT{disclose} operation of reporters, is to be visible only to certain
   clients, inside $S$ or outside of it, that represent stakeholder (such as auditors, and system
   managers) that have the need to get logging 
   reports, and have the right (however defined) to see them.

\item No other communication between components of MS and Internet cites
      outside of $S$ is permitted.

\item All other APIs provided by the components of MS are for internal use, and
      must be hidden from the rest of the system.
\end{itemize}    
\noindent

\noindent \textbf{Distributed Sandbox:}
Systems need sometimes to use untrusted code---untrusted because the code has
been developed by untrusted third parties, or if it is a newly developed part
of the system, which has not been fully
tested yet.  The safety of the host system requires  that the communication of
such code with the rest of the host system, and with the Internet outside of
the system, be carefully circumscribed and monitored.  More specifically, one
generally needs to impose constraints on which part of the host system can send
what kind of messages to the untrusted code. And, perhaps more importantly, one
need to constraint the kind of messages that the untrusted code can send to the
rest of the host system.

In local
systems this is accomplished by placing the untrusted code in what has come to
be known as a \emph{sandbox}, whose interaction with the host system is
strictly constraints. Such constraints are generally imposed either by the
language in which the host system and the untrusted code are written, or by the
 platform on which the system runs.

But when dealing with a distributed system, where the untrusted code can be
distributed as well, we do not have  a
programming language to rely on establishing a sandbox, as different system components can be written
in different languages. Nor can we rely on the platforms on which components
run, as  there may be many different platforms used by the system.

\s{Desirable Properties of  Modular Distributed Systems}\label{requirements}
As a first step towards the introduction of our model of a modular distributed
system (MDS) we describe in this section some of the properties that such a system
should satisfy.
These properties are realized by our basic model of MDS introduced in
\secRef{model}, which is a special case of the complete MDS model introduced in \secRef{advanced}.

We start  this section with a  schematic description of the structure of an MDS,
along with the terminology to be used throughout this paper. We then
characterize and motivate   the type of controls that one may want to impose
 over the flow of
 messages into  d-modules and from them. And we  conclude with some basic
properties that the  implementation of an MDS should satisfy.

\ss{The Basic Concepts and  Terminology of MDS:}
Schematically, an MDS  is a set of distributed components grouped into a
disjoint\footnote{This is a simplification,  d-modules do not have to be
disjoint under our complete MDS model.} set of \emph{d-modules}. 
(We will often refer to a d-module simply as a module, while 
the conventional  language-based  module will be referred to as
 as a \emph{traditional module}).  Every  d-module is composed of a set of one or more
 distributed components. But this is not a physical composition, but a logical
 one; having to do with the constraints on communication imposed on all
 component parts of a module.
The modules of a given MDS  are identified uniquely by their names, and each module can carry a
\emph{profile} consisting of an arbitrary list of \emph{labels}.
A module that contains just a single component is called a
 \emph{singleton}.

We use the following notation for various structural aspects of a given MDS $S$.
 Specific modules of $S$ are denoted  by capitalized symbols, such as $M$, $N$,
 or $M_i$;  and individual components of $S$  are  denoted by lower case symbols
 such as $c$, $d$, or $c_i$.
 A component $c$ that belongs to module $M$ is denoted by  $c^M$;
and the complement of module $M$ with respect to
the entire system $S$, is  denoted by $\overline{M}$.
Finally, actors operating
over the Internet  that do not belong to $S$ are called the
\emph{outside} of $S$, denoted by $outside(S)$.

\begin{figure}
\leavevmode
\epsfysize=3.0 in
\epsfxsize=4 in
\centerline{\epsffile{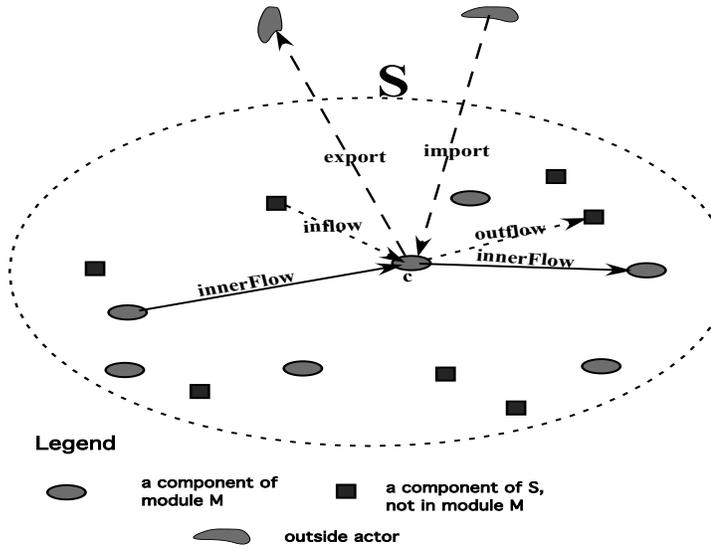}}
\caption{Various types of message flows with respect to a component $c$ of a
module $M$
\label{fig-module}}
\end{figure}

Due to the focus of MDS on the flow of messages into and out of modules, we
introduce the following notations about flow of messages with respect to
 a given module $M$ of $S$:
\begin{quote}
\noindent $\bullet inflow(M)$ is the receipt by a component of $M$ of a message sent
 from   somewhere in  $\overline{M}$;

\noindent $\bullet outflow(M)$ is the sending by a component of $M$ of a message addressed 
to some components in  $\overline{M}$;

\noindent $\bullet import(M)$ is the receipt by a component of $M$ of a message
sent  from somewhere in $outside(S)$;

\noindent  $\bullet export(M)$ is the sending  by a component of $M$ of a
message addressed  to  somewhere in  $outside(S)$.

\noindent $\bullet innerFlow(M)$ is the exchange of a message between the
components of $M$;

\end{quote}
\noindent
These types of flows are depicted in \figRef{fig-module}, which represents a
system $S$ whose components are partitioned into two types: (a) the components
that belong to a module $M$, depicted by ovals: and (b) the rest of the
components of $S$, depicted by squares. Also, shown, by irregular forms,  two
actors outside of $S$, somewhere over the Internet.  
Note  that the components of $M$, in this figure, are
intermixed with the other components of $S$, with no physical boundary that
marks them as belonging to a single module---we shall see in \secRef{model}
what makes a group of components into a module. 
The five types of flow defined above are depicted in this figure  by different
forms of arrows directed into 
a specific component $c$ of $M$, or away from it.

\ss{Control Over the Flow of Messages in an MDS}\label{control}
The main purpose of modularization of distributed systems is to  control
 the flow of messages into and out of the
various modules of a system. The following is a presentation of the controls we
require over the various types message-flow identified above. The definition of
each of this type of control is followed by a discussion written in italics,
and enclosed in a pair of curly brackets.

\begin{itemize}
\item \textbf{Inflow Control:} The imposition of constraints over which component of
      $M$  can receive which kind of messages from which other modules of $S$.

   \{{\em Discussion: This control is analogous to the conventional concept of
\emph{interface} of the traditional modules. But it is significantly more
general than the interface supported by most programming languages, in the
following sense: While the conventional interface makes certain methods of a
module visible everywhere in the system, the inflow control under MDS can make
certain APIs of a module accessible only to specified modules. The need for
such \underline{selective accessibility} has been demonstrated in
\secRef{motivation} by the monitoring service (MS) example, where the
\TT{disclose} operation  is to be accessible only to selected
modules. Indeed such selective accessibility is featured by  Eiffel
\cite{mey92-1}, where it is called ``selective export,'' but  most programming
languages do not support this capability.}\}

\item \textbf{Outflow Control:} The imposition of constraints over which
      component of $M$ can send which kind of messages to which other modules
      of $S$.

   \{{ \em Discussion: While inflow control---the only control provided by traditional
 modularization---is effective in protecting the internals of a given module
 $M$ from the rest of the system (namely from $\overline{M}$) it does not
 protect $\overline{M}$ from $M$. Because the inflow control leaves the
 internals of $M$ free to send arbitrary messages to 
components in  $\overline{M}$, which may change the behavior of the system in
unpredictable manner. This means, in particular, that the statement often made
about conventional modularization that the hidden internal of a module can be
changed freely without affecting the system---as long as the public interface is
implemented correctly---is not justified. 
This problem is particularly stark in the case of untrusted code discussed in
\secRef{motivation}, which needs to be placed in a sandbox to protect the
system against it.   But outflow control can address this problem quite
effectively.}\}

\item \textbf{Export and Import Control:} 
  The imposition of constraints over which component of
      $M$ can send (or receive) which kind of messages to (or from) 
 $outside(S)$.

   \{{\em Discussion: An example of the potential importance of such
control over the communication of a module with the outside of the system is
provided by the monitoring service (MS) example in \secRef{motivation}. The MS
group of components contains
sensitive information that needs to be shared with certain specific actors outside of the
system, but should not be leaked to anybody else over the Internet.

In conventional distributed systems such controls is provided by firewalls.
But exercising such controls under an MDS has several important advantage over
the firewall base control. In particular, a single set of
constraints, specified at a level of a module, can replace a whole collection
of distributed firewalls. }\}

\item \textbf{InnerFlow control:}   The imposition of constraints over the exchange of
      messages among the components of  module $M$.

   \{{\em Discussion:  Since the components belonging to a given module may be
constructed and  managed under different administrative domains, one may
want to impose constraints over their interaction. In the case of the
monitoring service of \secRef{motivation}, for example, one may want the
reporters, which are supposed to get their information from the analyzers, not
to have a direct access to the logs, lest they reveal raw data that needs to be
private. 
It should be pointed out that out of all the control types presented here,
this is probably the only one that is irrelevant to local systems.}\}

\end{itemize}
\noindent

\ss{Required Properties of the Implementation of MDS:}\label{properties}
We introduce and motivate below three key requirements from any implementation
 of the concept of MDS.
 
 \textbf{(R1) Independence of the Modular Structure of an MDS from the Code of System Components:}
 Modularization in distributed systems should be independent of the code of the
 components of the system, and of the programming languages in which they are
 written. The constraint on the flow of messages should be enforced by a
 middleware that views the components as communicating black boxes.
The reasons for this requirement are: (a) the code of different components may
be written in many different languages, and is not generally available for
inspection and for control by any single mechanism; and (b) the
 modularization-based constraints should be invariant of the evolution of the
 code. 

 \textbf{(R2) Modularity of the Specification of the Modular Structure of  an MDS:}
The specification of the constraints on the message flow into a given module,
and out of it, should be completely independent of  such specification
with respect to other modules.

 \textbf{(R3) Local Control:} The control over the flow of messages into or from a
given component $c$ of a given module $M$ should be carried out locally, at
$c$---or be carried out by a device dedicated to mediate the interaction of $c$
with others.
 In other words, the  control over flow into and out of a component $c^M$
 should not be done by a mediator that represent
 module $M$, mediating the message flow of all components of $M$. Because the
existence of   a mediator for each module would complicate the system, and
reduce the efficiency of mediation (This is  particularly true
 in the presence of hierarchy of
modules under the complete MDS model, to be discussed in \secRef{advanced}.)
 Also, this control should
not be carried out via a  system-wide mediator, because such a mediator would
constitute a single point of failure and an obvious target for attacks; and it
would be unscalable to boot. Moreover, a system-wide mediator would be
inconsistent with the R2 requirement above.

\s{On a Middleware Underlying Modular Distributed Systems}\label{middle}
The function of the middleware in the context of an MDS is to enable the
formulation of constraints over the messaging activities of distributed
components, and to carry out the enforcement of such constraints. Since
messages are the means for interaction between distributed components, we call
such constraints \emph{interaction-laws} or simply laws. And as we shall see in
\secRef{gist}, we will define a d-module to be the set of components whose
interactive activities are governed by a specific interaction-law---which makes
such laws central to our concept of MDS.
 The following are
some key requirements, regarding interaction-laws, posed by our concept of MDS.

\begin{enumerate}

\item  The enforcement of interaction-laws needs to be  \emph{decentralized},
       for several reasons: (a)  to  realize the requirement R2 and R3 above;
(b)  for the sake of scalability; and
(c) to avoid having a single point of failure, and a critical target for
    attacks.

\item  Interaction-laws need to be \emph{stateful}, so they can be
      sensitive to the history of interaction. (This aspect of interaction-laws
 is required for MDS for dealing with  crosscutting concerns,
that involve coordination between system
 components, as  hinted in \secRef{cross}.)

\item It should be possible to impose  multiple interaction-laws over
 a  single distributed system  in order to represent different modules,
 independently of each other. And these laws need to  \emph{interoperate}
 seamlessly, and  be
  organized  into what we call \emph{conformance
    hierarchy}. 
(This requirement will be explained and motivated in due course.)
\end{enumerate}
\noindent
We employ here a middleware called \emph{law governed interaction}
(LGI) that fulfills these requirements. 
Although LGI has been published extensively, there is no single
paper that describe all  the above mentioned properties of  it.
We outline, therefore, the LGI mechanism below, paying particular attention to
how it satisfies the above requirements.

\ss{The Law-Governed Interaction (LGI) Middleware---an Overview:}
 LGI is a middleware that can govern the interaction (via message
exchange) between  distributed \emph{actors}, by 
 enforcing an explicitly specified law---and possibly multiple laws---about 
such interaction. By the term ``actor'' we mean here
 an arbitrary autonomous process of computation, whose
structure and behavior is left unspecified, and is viewed as a black box by
LGI. Thus, an actor can be such things as a software component, an hardware
device, or a person communicating via a smart-phone.

We provide here a brief, and rather abstract, overview of LGI; focusing on the
properties of it which are the most relevant to this paper.  A more detailed
presentation of LGI, and a tutorial of it, can be found in its manual
\cite{min05-8}---which describes the release of an experimental implementation
of the main parts of LGI.  For additional
information and examples the reader is referred to a host of published
papers\footnote{These papers are available at
http://www.cs.rutgers.edu/~minsky/pubs.html.}, some of which will be cited
explicitly in due course.

The rest of this section is organized as follows.
We start, in \secRef{local}  with the local nature of the laws supported by
 LGI---a key characteristics of this middleware, which underlies the above
 mentioned properties. In \secRef{enforcement} we describe the law enforcement mechanism
 of LGI. And in \secRef{hierarchy}
we discuss the need for multitude of laws, and the hierarchical organization of
such laws.

\sss{The Local Nature of Interaction Laws Under LGI:}\label{local}
 Although the purpose of interaction laws is to govern the exchange of messages between
 different distributed actors,
 they do not do so directly under LGI. Rather, a law governs the \emph{interactive
activities} of any actors operating under it, in particular, by imposing
constraints on the messages that such an actor can send and receive. The ruling
of such a law for an \emph{interactive event} that may occur at any actor
$x$---such as the sending of a message by $x$, or 
 the  arrival of a message at it---can depend on the \emph{interactive state}, of the actor in
question, by which we mean,  some function of the history of interaction of
this actor with the rest of the system.  The   exchange of
messages between two actors, is therefore, governed by the laws under which
each of them operates, which may or may not be the same law, as we shall see.

This types of laws are \emph{local} in the sense that they can be enforced
locally, with no knowledge of, or dependency on, the interactive state of any
other actor of the system.  This means that the exchange of a message between
two actors requires their respective laws to be enforced separately, first at the
sender of the message, and then  at its receiver.
Such  decentralized  enforcement,  is, of course, very scalable, even
for highly stateful laws. 
Note that  without locality one would need to employ a central
mediator (or reference monitor) to mediate the interactions between all
actors, as it is done under conventional AC mechanisms, such as XACML. And interaction via 
 a central mediator is inherently unscalable for stateful policies.
 (This is the case even if the mediator is replicated,
as has been shown in  \cite{min99-5}.)

\sss{The Law Enforcement Mechanism of LGI:}\label{enforcement} The LGI
law-enforcement is carried out, broadly, as follows: For an actor $x$ to
conduct its interactive activities under a given law \EL, it needs to adopt a
software entity called \emph{controller} to serve as its private mediator,
subject to law \EL. The controller itself is generic, as it is able to operate
under any well formed LGI-law. So its adoption by $x$ involves loading the
specific law \EL\ into it, which creates a pair $\langle x, T^{\mathcal{L}}_{x}
\rangle$, where $ T^{\mathcal{L}}_{x}$ is the controller mediating the
interactive activities of $x$ under law \EL.  This pair is called an
\EL-\emph{agent}, since its interactive behavior, as seen by other actors, is
forced to conform to law \EL.
(Note that the act of adoption is one of the interactive
events of LGI, which signifies the birth of an agent, which may be followed
by an initialization procedure, as require by the law.)
The trustworthiness of the controllers is discussed in \cite{min05-8}. 

\begin{figure}
\leavevmode
\epsfysize=1.0 in
\epsfxsize=3.5 in
\centerline{\epsffile{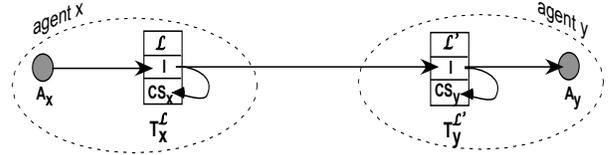}}
\caption{Interaction between a pair of  LGI-agents, mediated by a pair of
controllers under possibly different laws.
\label{fig-agent}}
\end{figure}

\figRef{fig-agent} depict the manner in which a pair of actors, operating via 
 different  controllers,, under possible different laws,
 interact with each other. Note that,
depending on security considerations,
the controllers may reside on the hosts of their respective actors, or they may
be anywhere over the Internet, but managed by a \emph{controller service} (CoS).

\noindent \textbf{On the Performance of LGI:}
An extensive study of the overhead incurred
due to the use of LGI has been reported in \cite{min99-5}, indicating that this
overhead is generally reasonably small. The evaluation of the law takes 
about   $50\mu s$, using a fairly standard PC, and for a relatively simple
 laws, of the kind one expect to use for  MDS.
This is  often negligible, particularly for communication over WAN.

\sss{About the Concept of \emph{Conformance  Hierarchy} of Laws:}\label{hierarchy}
As noted at the top of \secRef{middle}, an MDS-based system would need to be
governed by  multiple laws, 
organized  into, what we call, a
\emph{conformance hierarchy}, which we denote by $H$
In fact such organization of laws is required for many complex applications,
and it is supported by LGI.
Here we provide only a brief outline of this concept. For a formal and more detailed description
of conformance hierarchies of laws the reader is referred to \cite{min03-6}.

\p{The Structure of a Conformance  Hierarchy of Laws:}
A conformance  hierarchy  $H$ is a tree of laws rooted by a law  \law{R},
 such that every law \CAL{L'} in  $H$,
except its root law, is derived from its superior law \CAL{L} via a
derivation mechanism that ensures that \CAL{L'} \emph{conforms} to \CAL{L}---in
the sense to be described below---and
that this conformance is transitive.

Rather than using a uniform definition of conformance---such as requiring that
a subordinate law can deviate from its parent only by being more restrictive
than  it, which is a common
view concerning access control policies---we let each law
define what it means for its subordinates to conform to it.
This is done as follows:
 For a law to belong to a law hierarchy it needs to have two different parts,
 which we call the \emph{ground} part and the \emph{meta} part.
The ground part of a law \EL\ imposes constraints on 
 interactive behavior of the
actors operating directly under this law. While the meta part of \EL\ 
 circumscribes  the extent to which a  laws subordinate to \EL\ are allowed to
 deviate from the ground provisions of it.

 As a simple example, the
root law \law{R} may prohibit all interaction between components, while
enabling subordinate laws to permit any kind of interaction under their
purview. Alternatively, law \law{R} may permit all interaction, while enabling
subordinate laws to prohibit selected interactions.

This very flexible concept of conformance  is somewhat analogous to the manner in which state laws in the US conform
to the federal laws. And such conformance turns out to be very useful for the
governance of complex distributed systems as well, as we shall see in
\secRef{case}.

\p{The Formation of a  Conformance Hierarchy of Laws:} 
A conformance hierarchy  $H$ is formed incrementally via a   recursive process described
informally below.
First one creates the root law \law{R} of $H$.
Second,  given a law \EL\ already in $H$, one
defines a law  \CAL{L'}, subordinate to \EL, by means of 
 a law-like text called  \emph{delta}, denoted by
$\Delta($\CAL{L},\CAL{L'}$)$, which specifies the intended differences between
 \CAL{L'} and \EL. This is done in a manner that ensures that
 law \CAL{L'} conforms to its
superior law \EL---for a formal model of this derivation see  \cite{min03-6}.

\s{A Basic Model of MDS}\label{model}
 An  MDS, according to this basic model, is
a triple    $\langle C, H, E\rangle,$ where  $C$ is a set of distributed
components that populate the system;   $H$ 
is a  conformance hierarchy
 of laws (also called the
 \emph{law ensemble})  that  defines the modular structure of a system; and $E$ is the LGI-based mechanism that
 enforces the laws of $H$, thus establishing this structure.

The only assumption made by this model about the components of $C$ is  that
they all  communicate via LGI, subject to laws  in $H$.
 We will justify this assumption in \secRef{semi-enforcement}, under
certain conditions, and will qualify it when these conditions are not
satisfied. The law ensemble $H$ is bi-level under this basic model, but as we
shall see in \secRef{advanced}, it can be extended to arbitrary depth. And the enforcement mechanism $E$ is a middleware
consisting of LGI controllers that mediate the interaction between the components of $S$,
subject to the various laws in $H$.

The rest of this section is organized as follows: 
we start with the gist of this model, which explains how modules
of MDS are to be defined, and how the overall modular structure of a system is
established.   \secRef{case} is a rather generic case study that illustrates 
 how the  modular structure of an MDS is defined.   \secRef{semi-enforcement}  elaborates
on our assumption that the laws in $H$ are strictly enforced; and
\secRef{scratch} discusses the construction of an MDS.

\ss{The Gist of MDS:}\label{gist}
The key concept of MDS is that of a distributed module, defined below:

\begin{figure}
\leavevmode
\epsfysize=1.5 in
\epsfxsize=3.5 in
\centerline{\epsffile{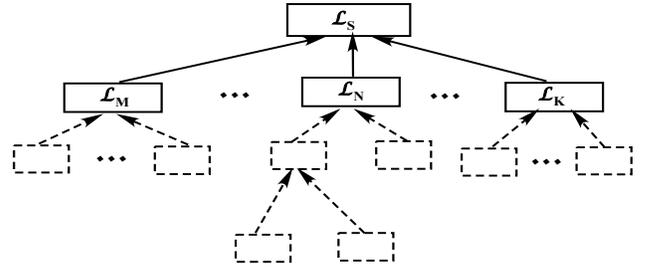}}
\caption{An Hierarchical Law-Ensemble of an MDS
\label{fig-laws}}
\end{figure}

\begin{definition}
A distributed module $M$ of an MDS $S$ is a set of components of $S$ 
that communicate with others subject to an interaction-law \law{M} that belong to the lawp
ensemble $H$ of $S$.
\end{definition}
\noindent
This definition of a module via a law under which its components communicate,
has several important implications. First, it provides an unambiguous
definition of the composition of modules---namely, the set of modules that
operate under this law. Second,
 law \law{M} is
 in a position to impose constraints on the flow of messages
into and from each component $c$ of $M$. These
 constraints are enforced locally
 at $c$; due to the decentralized nature of LGI, and according to requirement R3 of
\secRef{properties}. 
Third, this definition provide great flexibility to this model, facilitating such
things  as and the ability of modules to
overlap---which is a feature  of the complete MDS model.
(Note also that under the 
LGI-terminology,  a module $M$, thus defined, is precisely   the \law{M}-community.)

Such module-laws can be defined independently of each other, and according
to  requirement R2 of
\secRef{properties}. But there must be some similarities between these laws,
for the interaction between modules to be coherent. For example, there must be
some common way for interacting components to identify the names of the 
 modules to which
they belong. And one may want some constraints on
the flow of messages to apply to all  modules of a given system.
  We refer to such commonalities as 
  \emph{regularities} of the system.

Under this model regularities are
established, by having all module-laws, such as \law{M} above,
subordinate to a single law, which serves as the root law of an hierarchical
law ensemble  $H$. Such a law ensemble is 
 depicted in \figRef{fig-laws}. It is a bi-level\footnote{Note that this figure
 contains  laws  situated below the second level---they are depicted by dashed
 lines. Such laws are possible under the  complete model of MDS, which
 supports law hierarchies of
 arbitrary depth.} hierarchy, whose root law is
 denoted but \law{S}, where $S$ is the name of the system at hand.
The individual laws in this hierarchy are maintained by what is called a
\emph{law server}, which ensures that the name given to these laws
are unique.

\begin{figure}
\leavevmode
\epsfysize=3.5 in
\epsfxsize=3.5 in
\centerline{\epsffile{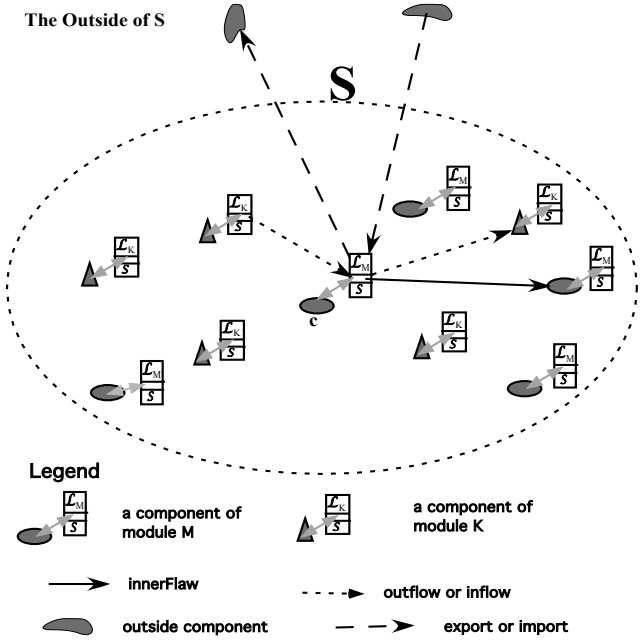}}
\caption{A Schematic Depiction of an MDS
\label{fig-arch}}
\end{figure}

Given such a law ensemble designed for an MDS system $S$, and a set $C$ of
components, the system itself is constructed by having each
component $c$ in $C$ that is intended to belong to module $M$,
adopt an LGI controller with
 law \law{M} that  defines this module.
 The interactive activities of $c$ would
then be governed by law \law{M} of the module to which it belongs; and,
indirectly,  by the
root law \law{S} to which law \law{M} conforms due to the hierarchical
structure of $H$. A system of this kind is depicted schematically in
 \figRef{fig-arch}---which is an elaboration of \figRef{fig-module}. It shows two modules, called $M$ and $K$, whose
 components are dispersed over the Internet (the components belonging to module
 $M$ are
 represented in this figure by ovals, and those belonging to  $K$ are
 represented by triangles.). There is no physical boundary
 enclosing the members of a module; their membership is defined by the
 laws---\law{M} and \law{K}, respectively---under which they communicate.
This figure also depicts the various flow of messages into and from components
$c$ of module $M$.

It is worth pointing out here
 that this basic model provides no control over which
components can belong to which modules, but such control is provided by the
complete MDS model, introduced in \secRef{advanced}.

An important consequence of this model  is that  \emph{modules have no
physical embodiment}.
Rather, a modules is defined as a set of of components whose
interactive activity is governed by the same law. And as required in
\secRef{properties}, the enforcement of the law of a module is carried out by
the private controller  associated with each component.
The only physical manifestation of a module is its law, which is
maintained at the
controllers of the  various components operating under it, and at the
law-server.

Finally, it should be pointed out that a singleton module $N$, i.e., a module
that contains a single component $c$ has advantages over a having $c$ not be
contained in this module. Because the law \law{N} of $N$ can provides
guarantees to the rest of the system that cannot be obtained otherwise.
This, by limiting the outflow and the export and import that can be carried out
by $N$.

\ss{Establishing the Modular Structure of an MDS---a Generic Case
Study:}\label{case} To illustrate how a modular structure of an MDS can be
defined by its law ensemble $H$---and  enforced by $E$---we describe here the
law ensemble of a more or less generic case study.

 The various laws in $H$ are described here broadly and informally. 
A reader who wishes to see how such laws are actually written under LGI is
referred to \cite{min03-6}, where a fairly sophisticated hierarchical ensemble
of laws is introduced in details. Here are some comments
 about our
informal description of  $H$. 
A typical  law consists of several rules, which represent different
provisions regarding different aspects of the modular structure.
  And all but the root
law of this hierarchy are represented by their deltas, which specify the
 differences between the law at hand and its superior.
   We employ here the following
convention about the meta\footnote{The concepts of \emph{delta} and of
\emph{meta} part of a law have been introduced in \secRef{hierarchy}.} part of any given law \CAL{L} in the hierarchy: (a)
if \CAL{L} has a rule that addresses a certain aspect of messaging activity
of a component subject to this law---such as the sending of a certain type of
message---then this rule  cannot be deviated from by subordinate (and transitively subordinate)
laws of \CAL{L}, unless such deviation is explicitly permitted by the meta part
of \CAL{L}---such meta permissions are denoted by bracketed text in bold  
italic font; and (b) if \CAL{L} has nothing to say about certain aspect of
interaction, then subordinate laws have the freedom of legislation about it.
We start
 with law \law{S}, and then  continue with an example of a module-law
 subordinate  to it.

\sss{The Root Law \law{S} of $H$:}\label{global}
Due to the conformance nature of the hierarchical law ensemble $H$, its root
law \law{S} has  dominion over all the laws in $H$.
Consequently, this law governs  all
the messaging activities of components of the system at hand\footnote{This is
true under the assumption made at the top  of  \secRef{model}, that all message
exchange in the system  is done
under laws in $H$---but see \secRef{semi-enforcement} for an elaboration of
this assumption.}. Some of the rules of this law are stated categorically, not
allowing any deviation from them  by the  subordinate module-laws---thus
establishing system regularities. Other rules in this law, which permit deviations
from them by module laws, can be viewed as establishing defaults.
Note that most of the rules below are followed 
by  a discussion---in italics enclosed in curly brackets---that elaborates
 on the rule, providing 
some  clarification of it and  motivation for it.

\begin{enumerate} 

\item \textbf{Initialization:} The name $M$ of the module-law \law{M} that
      has been adopted by a component $c$ is to be 
  stored in the state of its
adopted  LGI-controller.

\textbf{\emph{[But the module-law itself may add conditions to this
rule, and may require various  operations to be carried out upon adoption,
perhaps for adding information that further identifies the module.]}}

   \{\emph{Discussion:  Having the unique name of a module, represented at every
   component of it, is essential for flow control as is explained in the
   context of \ruleRef{R-id} below.} \} 
\label{R-init}

\item \textbf{Sender identification:}
 Every message sent is to be concatenated with the name and profile
 of the module to which
 the sender-component belongs.  This identifier  is to be
stripped from the message, before  it is delivered to the target component
itself. 

   \{\emph{Discussion:  This sender identification is needed by the receiver of
   the message, 
   for it to be able to determine if this message satisfies its own
   inflow-control rules, and thus aught to be accepted---the profile can be
   specified by the module law, as discussed in \secRef{module-law}.
  Note that this identification is
intended mainly for the controller of the receiver, and  it is stripped
from the message, before it is delivered to the component itself.
This,  in order to
accommodate legacy components,  which would not know what to do with the extra
  information added to messages by the law. It is worth pointing out here that
  a component written with knowledge of
 this law can get the identifying information by explicitly asking its
 controller to
 disclose it. But such disclosure needs to be permitted by the law under which
 one operates.}\}
\label{R-id}

\item \textbf{Default inflow control:} All inflows of messages are prohibited,
\textbf{\emph{[unless permitted by the  subordinate module-law in question]}}
\label{inflow}

\item \textbf{Default outflow control:}
All outflows of messages are permitted,
\textbf{\emph{[unless prohibited by the  subordinate module-law in question]}}
\label{outflow}

\item \textbf{Default export/import control:}
All exports and imports are prohibited,
\textbf{\emph{[unless permitted by the  subordinate module-law in question]}}.

\item \textbf{Default innerFlow control:}
All message exchanges between member of a module are permitted,
\textbf{\emph{[unless prohibited by the  subordinate module-law in question]}}
\label{innerflow}

   \{\emph{Discussion: The above four rules establish defaults controls over
 inflow,   outflow,  export/import and innerFlow. But they  allow individual
 module-laws to override these  defaults.
The rationale of these particular defaults is as follows: Regarding inflow control,  assuming that relatively few inflows of messages into a module 
 would end up being allowed, \ruleRef{inflow} prohibit them all,   as a default.
  But it enables individual module-laws to permit  arbitrary inflows.
The other three default rules above can be justified  by similar consideration.
 But one can, of course, design different kinds of default rules.}\}
\label{export}
\end{enumerate}

\sss{Module-Laws:}\label{module-law}
We discuss here  a single  module-law    \law{M} of some  module $M$.
This law  is a subordinate to the system-law
\law{S}, which is  derived from \law{S}
 via a delta $\Delta($\law{S},\law{M}).
 Below is an informal description of  a typical  such  delta,
 distinguishing between two aspects of it: (a) initialization,
 to be done upon the adoption of a
 controller with \law{M}; and (b) imposition of control  over the flow
 of messages into $M$ and from it.

 \textbf{Initialization:} Law \law{M}
may mandate that   a given set  of labels would  be added to
the state of each component operating under this law, as the \emph{profile} of
the module.
Recall, that as required  by the system-law \law{S}, 
this profile would be appended to every message sent by every component of
module $M$, in order to identify it.

 \textbf{Flow Control:} Recall that the system-law \law{S} establishes defaults for all four
types of flows of messages we identified, allowing module-laws to change these
 defaults  arbitrarily. In particular, law \law{s} prohibited all inflows.
But law \law{M} can permit specific types of inflows, as follows:
 It can permit certain types of messages  from anywhere in the system---which is
equivalent to conventional concept of an interface of a module.
Or,  it can
permit such messages to come from one or  several modules, which can be specified
by their names, or by their profiles. The defaults of other types of message
flow can be changed in a similar way.

\ss{About the Enforcement of  the law-hierarchy of an MDS:}\label{semi-enforcement} 
We have assumed above that  all  components of $S$ satisfies the following conditions: (a) they
   communicate via LGI, and (b)  they operate subject to laws in $H$. 
Under this twofold assumption the law ensemble
 $H$ is clearly enforced by the trusted LGI controllers. 
But how can one ensure that this  assumption is  valid?

First note that if part (a) of this assumption is satisfied, than its part (b)
can be established simply by having the system-law \law{S} require that messages
can be received only if they are sent by component operating under laws that
are subordinate to \law{S}---that is,  laws in $H$.
So, if a component chooses to operate under LGI law that does not belong to $H$,
it will not be able to communicate with any components operating under $H$, and
could than be considered not to belong to $S$.
But part (a) of this assumption is more problematic, because we may  have
no control over how a set of distributed  software components
communicate with each other.

 Assumption (a) can be forced  to be  satisfied 
if all components of $S$ are on a single Intranet, or on  a set of Intranets
 managed under a single administrative domains,
  where one has control over the
local  network (or networks) and its firewalls. This   has been demonstrated  in \cite{he05-1}.

But even when   of a given system is dispersed throughout the
Internet, its components  may often be \emph{virtually compelled} to operate
 under some law in $H$, or even under a particular law \CAL{L} in $H$.
  Broadly, this is the case for a component that needs to
communicate with other components,  which  require their  interlocutors to  operate under a
given law \CAL{L}, or under any subordinate law to it. This is so,
basically, because one can detect whether  its interlocutor
operates under LGI, and can identify the law under which it operates---this is
one of the essential features of LGI, called ``law-based trust'' \cite{min12-2}.

\ss{The Construction of an MDS}\label{scratch}
The construction of an MDS   $\langle C, H, E\rangle$ from scratch is fairly
straightforward: One first defines the  the modular structure of the system
via the law hierarchy $H$,
and then associates components with the various modules, by having them adopts
the suitable laws in $H$. Of course, it may not be  that simple, because one often
 needs
to change the modular structure, incrementally, during the construction process.
We will address this issue when discussing the evolution of an MDS, in
\secRef{future}. The conversion of a legacy system to an MDS is an open
problem,  also discussed
in \secRef{future}.

\s{The Complete MDS Model}\label{advanced}
 Our complete MDS model provides the
following capabilities omitted,  for simplicity,  from the basic model:
(1) controlling  and reporting  the  membership of modules;
(2) virtual nesting of modules;
(3) allowing modules to overlap;
and (4) implementing crosscutting modules.
We introduce these capabilities below, including their motivation and
implementation. 
It is worth pointing out that  these capabilities do not require any change in
the underlying structure of the basic model, or in the
present state of LGI. 

\ss{Controlling and Recording the Membership of Modules:} Note that the basic
MDS model provided no means for imposing constraints on which components can
belong to a given module, and for recording the actual membership of a given
module, at a given moment in time.  This is unacceptable for many reasons. In
particular, not being able to constrain the membership of a module may pose
serious security risks, as it may allow a rogue component to enter some
sensitive module, such as one that implements the monitoring service of
\secRef{motivation}. And not being able to record the membership of module
would make the management of a system very difficult.  However, these
capabilities can be provided under the complete model of MDS, in several
ways---such as described below.

 \textbf{Controlling the Membership of a Module:}
Suppose  that each component of a given system $S$ has a private key, and that
there is   a certification authority (CA)  that
 provides each component with a digital certificate that identifies its
unique name (unique with respect to  system $S$), and
the module (or modules)  to which it is allowed to  belong.
Under this condition, the module-law of a given module $M$ can be written to
require such a certificate upon the adoption of this law, and to refuse to
be adopted if the right certificate is not presented. In fact, such control
over membership may be made into a system regularity, if the above is done
 not by individual
component-laws but in the system-law \law{S}.

 \textbf{Recording the Membership of Modules:}
Suppose that a given system employs a
 server---called \emph{system registry}---to maintain the list of
system components, each identified by its name, IP
address,  and by the  module (or modules) to which it belong. This server can
be fed with the required information, by having the system-law \law{S} mandate that whenever a
components adopts a controller under a given component-law, a message will be
sent automatically to the registry, with its identification, and with the
name of the module it operating from.

\ss{Virtual Nesting of  Modules Under MDS:}\label{nesting}
Modules can be nested under MDS, simply by 
extending the depth of the conformance  hierarchy of laws $H$ to more then 2,
as  depicted by dashed
 lines in \figRef{fig-laws}.
 Of course, we
 do not mean physical nesting, but a logical one, in terms of the constraints imposed on the
messages that can flow into and out of the components of the nested modules.

For example, consider a module $M$ defined by  law \law{M}, anywhere in  $H$
below \law{S},
and a module $M'$ defined by  law 
\law{M'}
subordinate to \law{M}. Given the nature of conformance under LGI,
the 
components of  $M'$ operate under the same restrictions on  their
communication, as the components of $M$ itself---unless
\law{M} permit its subordinates laws to deviate from it in some sense.
Suppose, in particular, that
\law{M} permit its subordinate laws to strengthen (but not weaken)
 its own constraints on
communication. Then \law{M'} may impose additional constraints on how its
own components interact with the rest of the system.
 The potential benefits of such an hierarchical organization of
modules,  particularly for very large and complex
systems, seems self evident.

It is worth pointing out that this concept of nesting enables us to 
 view the entire system as a single \emph{universal module}, defined by
the system-law \law{S} of $H$, which has the other modules nested within it.

\ss{The Ability of Modules to Overlap:}
Modules can overlap in the sense that a single component may belong to several
modules.  This can be done by a component simply by adopting several LGI
controllers, under several different module-laws.
This basic capability of LGI  may be useful for several reasons, the following
is  of them; another reason is presented in \secRef{cross}, below.

Consider  a web server that provides several
different services. Such services may need to belong to different modules,
 because they may need to interact with different
set of components, and may require different inflow,  outflow and export/import
 controls. This is possible to do with the  overlap capability.
  
\ss{Crosscutting Modules:}\label{cross}
 The ability of components of an MDS to belong to
several modules enables the implementation of a distributed version of
crosscutting modules, thus extending to open distributed systems an important
 concept introduced under \emph{aspect oriented programming} AOP
\cite{kic05-1}. We illustrate this capability of MDS with the following example.

Consider an MDS-based system $S$ in which several components that belong to several
different modules engage in sending purchase orders (POs) to Internet cites
outside of
$S$. And suppose that the sending of POs is required to comply with  a given system
wide protocol that involves an approval workflow and logging of the POs
themselves.
Normally one would have to program this  protocol into every
component that issues POs---which is laborious and error prone process. But we can  ensure that this protocol is observed
for sending POs, by anybody who does it, simply by localizing the sending of 
POs in a single,
crosscutting module, say $P$. This can be done as follows.

First, we write a module-law \law{P} that enforces the required protocol, for
every PO being sent.
 Second, we have to write the root law \law{S} so that it
prohibits the sending of POs from anywhere,  except from module $P$. This
would force  components that need to send POs to enter module $P$---in addition
to their native module---by adopting
its law \law{P}.

\textbf{Related Work:} It should be pointed out that this is not the first implementation 
of  crosscutting concerns in
distributed system.
In particular, the  DaDO system 
 \cite{woh03-1} implemented such concerns by planting appropriate mediators,
 called \emph{adaplets}, in the code of the relevant components. Adaplets are
 analogous to our controllers with specific laws. But planting them into the 
 code of the various components, is overly laborious and unsafe.
In any case, unlike in DaDO, the crosscutting concerns under MDS are simply an
integral part of the more general concept of modularization.

\s{Some Open Problems}\label{future}
Although the MDS model can be used in its present form, it
raises  several issues that  require
further research and experimentation. Two such open issues are presented below.

\p{(1) The Evolution of an MDS:}
 Both the  base system of an MDS $S = \langle C, H, E\rangle$,
 (i.e., the code of the set of components $C$ of $S$), and its modular structure
 defined by $H$, are bound to evolve.
 The evolution of the code of the system presents no new problems under
 MDS. Quite the contrary, such evolution becomes safer, because no changes in the code can violate the constraints
 imposed by the modular structure $H$.
 This is, in fact,
 one of the most significant and advantageous aspects of the concept of MDS. 
However, the evolution of $H$, which defines the modular structure of $S$,
confronts the following  presents difficulties:
 (1) the potentially disruptive effect that  a change of $H$
may have on the system governed by it; and (2) the difficulties in carrying out 
 changes of non-leaf laws that belong to the law-hierarchy $H$ (note that
 changes  in leaf laws of $H$ present no difficulty.)

\p{(2) Evaluation of the Potential  Impact of MDS on the Engineering of Distributed Systems:}
The model of MDS introduce here has been tested experimentally on a couple of
 small systems, as a proof of
concept.  But the real usefulness of MDS, and its potential impact on the
engineering of  distributed system, cannot be validated
without applying it 
 to large, complex and open  distributed system. This calls for three kinds of
 experiments: (a) constructing a complex MDS from scratch; (b) converting  a
 large and complex 
 legacy system  to an MDS; and (c) subjecting one of such systems to a process
 of evolution. Such experiments are yet to be done.

 \s{Conclusion}\label{conclusion} 

We have introduced the concept of modular distributed system, composed 
of what we have  called distributed modules. Each such module
is comprised of a set of one or more distributed components, whose communication
with other modules of that system, and with the outside, is tightly
circumscribed. In other words, an MDS erects
 virtual, selectively permeable, boundaries 
between groups of components dispersed over the Internet.
This modular structure is established via a decentralized middleware, and is,
therefore very scalable.

This concept is inspired by modularization in local (non-distributed)
systems, and it is similar to it in that it provides for hiding.
But MDS  has  several important features that are rarely, if
 ever, supported by conventional modularization. 
These include
constraints on the ability of the body of a module to send messages to other
modules, or to the outside of the system; the ability of different modules to
overlap; and the ability to construct crosscutting modules. 

Although the full evaluation of the potential impact of the  MDS model has not been done yet, this impact is expected to be substantial,
  particularly once the open problems presented in \secRef{future} are solved.

\bibliography{biblio1}

\end{document}